\documentclass[aps,pre,showkeys,showpacs,superscriptaddress,
amssymb,amsfonts,twoside,twocolumn,floatfix]%
{revtex4} 
 
\preprint{bunch7}

\usepackage[dvips]{graphicx} 
\usepackage{psfrag} 
\usepackage{exscale} 
\usepackage{rotating} 
\usepackage{amsmath} 
\usepackage{bm}
\usepackage[latin1]{inputenc} 

\usepackage{ulem} 
\def\emph#1{{\it#1}} 

\renewcommand{\approx}{\simeq}

\def\rmd{\ensuremath\mathrm{d}} 
\def\kB{\ensuremath{k_B}} 
 
\def\growth{\ensuremath{\nu}} 
\def\nrSteps{\ensuremath{N}} 
\def\Prob{\ensuremath{\mathcal{P}}} 
 
\def\Eimp{\ensuremath{E_{\textrm{im}}}} 
\def\Jimp{\ensuremath{J_{\textrm{im}}}} 
\def\Rimp{\ensuremath{\rho}} 
\def\inFlux{\ensuremath{\Phi}} 
\def\probSingle{\ensuremath{p}} 
\def\dj{\ensuremath{d}} 
\def\xj{\ensuremath{d}} 
 
\def\rmd{\ensuremath{\mathrm{d}}} 
 
\def\ie{\emph{i.e.,~}} 
\def\cf{\emph{cf}~} 
 
\def\eq#1{(\ref{eq:#1})} 
\def\Eq#1{Eq.~(\ref{eq:#1})} 
 
\def\Fig#1{Fig.~\ref{fig:#1}} 
\def\Tab#1{Table~\ref{tab:#1}}

\begin{document} 
 
\title{Impurity-induced step interactions: a kinetic Monte-Carlo study} 
\date{\today} 
 
\author{J\"urgen Vollmer} 
\email{juergen.vollmer@ds.mpg.de} %
\affiliation{Fachbereich Physik,  
  Philipps Universit\"at,  
  Renthof 6,  
  35032  Marburg,  
  Germany} %
\affiliation{Dept.~Dynamics of Complex Fluids,  
  Max Planck Institute for Dynamics and Self-Organization,  
  Bunsenstr.~10,  
  D-37073 Göttingen, 
  Germany} 
 
\author{J\'ozsef Heged\"us} 
\affiliation{Fachbereich Physik,  
  Philipps Universit\"at,  
  Renthof 6,  
  35032  Marburg,  
  Germany} %
\affiliation{Department of Chemistry, 
  University of Cambridge, 
  Lensfield Road,  
  CB2 1EW Cambridge, 
  UK} 
 
\author{Frank Grosse} 
\affiliation{Institut f\"ur Physik,  
  Humboldt-Universit\"at zu Berlin,  
  Newtonstr.~15,  
  12489 Berlin,  
  Germany} %
 
\author{Joachim Krug} 
\affiliation{Institut f\"ur Theoretische Physik,  
  Universit\"at zu K\"oln, 
  Zülpicher Str.77,   
  50937 Köln,   
  Germany}

\begin{abstract}  
  A one-dimensional continuum description of growth on vicinal 
  surfaces in the presence of immobile impurities predicts that the 
  impurities can induce step bunching when they suppress the diffusion 
  of adatoms on the surface. 
  In the present communication we verify this prediction by kinetic 
  Monte-Carlo simulations of a two-dimensional solid-on-solid model. 
  We identify the conditions where quasi one-dimensional step flow is 
  stable against island formation or step meandering, and analyse in 
  detail the statistics of the impurity concentration profile. The 
  sign and strength of the impurity-induced step interactions is 
  determined by monitoring the motion of pairs of steps. Assemblies 
  containing up to 20 steps turn out to be unstable towards the 
  emission of single steps. This behavior is traced back to the small 
  value of the effective, impurity-induced attachment asymmetry for 
  adatoms. An analytic estimate for the critical number of steps 
  needed to stabilize a bunch is derived and confirmed by simulations 
  of a one-dimensional model. 
\end{abstract} 
 
 
 
\maketitle

\section{Introduction} 
 
Impurities and adsorbates affect the growth of crystals and thin films 
in a variety of ways. Small amounts of CO strongly enhance the 
nucleation density in homoepitaxial growth of Pt on Pt(111) and 
reverse the orientation of the resulting triangular islands 
\cite{Kalff98,Michely04}; a floating monolayer of a suitably chosen 
surfactant species induces layer-by-layer growth in many growth 
systems \cite{Michely04,Kandel00}; and the adsorption of antifreeze 
proteins on the growth surface prevents the formation of macroscopic 
ice crystals in the blood of fish living in polar waters 
\cite{Sander04}. 
 
On a vicinal surface growing by step propagation, impurities are 
generally expected to slow down the steps by pinning. This can lead to 
the formation of step bunches \cite{Cabrera58,vdEerden86,Kandel94}. A 
different mechanism for impurity-induced step bunching not related to 
step pinning was recently proposed in the context of SiC growth on 
Si(100) where C plays the role of a codeposited impurity 
\cite{Croke00,Krug02} (see also \cite{Amaral97}). Due to the motion of 
the steps, at any given time different parts of a terrace have been 
exposed to the impurity flux for different durations, which leads to a 
gradient in the impurity concentration directed towards the ascending 
step. The coupling of the impurity concentration profile to the 
diffusion of the growth units on the terrace may destabilise the 
equidistant step train. Originally \cite{Croke00} a lower binding 
energy to the impurities was suggested and confirmed by kinetic Monte 
Carlo (KMC) simulations as a possible physical source for the 
experimentally observed behavior.  Another scenario was discussed in 
\cite{Krug02}: Impurities slow down the adatom diffusion without 
affecting adatom binding energies (random barriers \cite{Haus87}) 
which also leads to instability. 
 
The linear stability analysis of \cite{Krug02} was based on a 
one-dimensional model of straight steps. The impurity and adatom 
concentration fields were treated in a continuum approximation and 
assumed to take on their stationary profiles instantaneously on the 
time scale of step motion.  In the present communication we revisit 
the problem within a fully microscopic KMC simulation, taking explicit 
account of non-stationarity and fluctuations. 

We identify a range of parameters where the assumption 
of a one-dimensional array of straight steps is applicable.  
We then consider systems of two, three and more steps with periodic 
boundary conditions in order to follow the loss of stability of the 
equidistant arrangement, and characterise the long-term evolution of 
the system. The basic stability properties predicted by the continuum 
theory are confirmed. Simulations with two steps show bound pairs and 
equidistant steps with only small fluctuations of the terrace width in 
the appropriate parameter regimes.  
However, we also find that the impurity-induced step interactions are 
unable to stabilise larger assemblies of steps.  Simulations with 3, 
4, 8, and 20 steps approach highly dynamic states with many closely 
adjacent pairs of steps that frequently exchange partners. 
This behavior can be explained within the framework of a 
deterministic step-dynamical model \cite{Popkov06}: in spite of 
attractive interaction between the steps, bunches that contain less 
than a critical number of steps decay by step emission.

The paper is organized as follows. In the next section the  
KMC model is introduced and the fundamental growth modes 
are identified as a function of the system parameters. 
Section \ref{Sec:Imp} contains a detailed analysis of the 
spatial distribution of impurities on the terraces, focusing 
in particular on the fluctuations around the mean impurity  
concentration gradient. The dynamics of step pairs, triplets and 
bunches is described and analyzed in Sec.\ref{Sec:Dyn}, and  
conclusions are given in Sec.\ref{Sec:Con}.

\begin{table} 
\begin{tabular}{l@{\qquad}l} 
\hline 
\textbf{parameter} & \textbf{description} 
\\[2mm] 
$J = \exp \left(\frac{E_b}{k_B T}\right)$ 
&  suppression of diffusion by bonds 
\\[2mm] 
$\Jimp = \exp\left(\frac{\Eimp}{k_B T}\right)$  
&  change of diffusion by impurities 
\\[2mm] 
$\growth = g/\inFlux$   
& hopping rate / deposition rate 
\\[1mm] 
$\Rimp$  
& fraction of impurities in deposited atoms 
\\[2mm] 
\hline 
\end{tabular} 
\caption{Dimensionless parameters characterising the diffusion of adatoms and  
surface growth.  
\label{tab:parameter} 
} 
\end{table} 
 
\section{Microscopic growth model}

We consider an SOS system with a simple cubic lattice, where the 
surface has an extension $L_i\times L_j$. There are periodic boundary 
conditions along $i$, and Lees-Edwards boundary conditions along $j$. 
The latter are also periodic on the surface, but they induce a change 
of height of \nrSteps\ steps when transversing the system in 
$j$-direction. By this topological constraint one enforces the 
existence of \nrSteps\ steps on the surface which are aligned 
parallel to $i$. Depending on the growth parameters these steps may be 
fairly straight, they may meander, or there may be additional islands 
on the surface. 
 
The kinetic Monte Carlo simulation is constructed to be close to the 
continuum description \cite{Krug02}. In the following the technical 
details are described with special emphasis on the differences to 
previous simulations \cite{Croke00}.  The transition rate of a 
thermally activated hopping process from site A to the neighboring 
site B is given in transition state theory by \cite{Michely04} 
\begin{equation} 
\Gamma^{\left({A \rightarrow B}\right)} = \Gamma_0^{ \left({A \rightarrow B}\right)} \exp\left(-\frac{E_A^{\left({A \rightarrow B}\right)}}{k_B T}\right). 
\label{EQ:01} 
\end{equation} 
The preexponential factor $\Gamma_0$ is taken usually to be $10^{13}$ s$^{-1}$, but other values are possible also~\cite{Grosse02}. The activation energy 
\begin{equation} 
E_A^{\left({A \rightarrow B}\right)} = E_T^{\left({A \rightarrow B}\right)} - E_B^{\left(A \right)} 
\label{EQ:02} 
\end{equation} 
is given by the difference of transition energy $E_T$ and binding 
energy $E_B$. $E_T$ may depend on initial and final state, whereas 
$E_B$ only depends on the initial state. From now on these 
dependencies are suppressed.  In the case of a simple cubic lattice 
usually the binding energy is described by a next neighbor counting 
model \cite{Shitara92} 
\begin{equation} 
E_B = - (E_S + n E_b) 
\label{EQ:03} 
\end{equation} 
with $n$ being the number of in-plane next neighbors.  In the present 
work the impurities are taken to influence the transition energy $E_T$ only, 
\begin{equation} 
E_T = E_t + \Eimp 
\label{EQ:04} 
\end{equation} 
with $E_t$ being the transition energy of free adatoms.  Previous 
simulations~\cite{Croke00} had only considered the influence on the 
binding energy $E_B$, which leads to a strong step bunching effect. 
Inserting Eq.~(\ref{EQ:03}) and (\ref{EQ:04}) into Eq.~(\ref{EQ:01}) 
results in the hopping rate of an adatom residing on an impurity 
\begin{equation} 
\Gamma = g \frac{1}{J_{im} J^n} 
\label{EQ:05} 
\end{equation} 
with 
\begin{eqnarray} 
g = \Gamma_0 \exp\left({ - \frac{E_t + E_S}{k_B T}}\right).  
\label{EQ:06} 
\end{eqnarray} 
The factors $J$ and $\Jimp$ are defined in Tab.~\ref{tab:parameter}. 
Therefore, free adatoms on the surface perform a random walk with 
hopping rate $g$.  Their hopping rate is 
modified by the existence of neighbors and impurities. 
 
In the simulation adatoms are added randomly to the surface with a 
flux of $\inFlux$ atoms per lattice site, 
and the ratio of diffusion and incoming flux will 
henceforth be characterized by $\growth \equiv g/\inFlux$.  An adatom 
on the surface will perform on average 
$\growth /( L_i L_j)$
\emph{free} steps before another atom is added 
anywhere on the surface. 
 
Impurities disappear from the surface by burying them under normal 
atoms, and they are created by a flux onto the surface. A fraction of 
$\Rimp$ of impinging atoms are impurities. When they hit the surface 
they \emph{immediately} exchange position with a ``normal'' atom in 
the surface. Subsequently, there is a new impurity in the surface, and 
an additional adatom diffusing on the surface. 
 
\begin{figure} 
\includegraphics{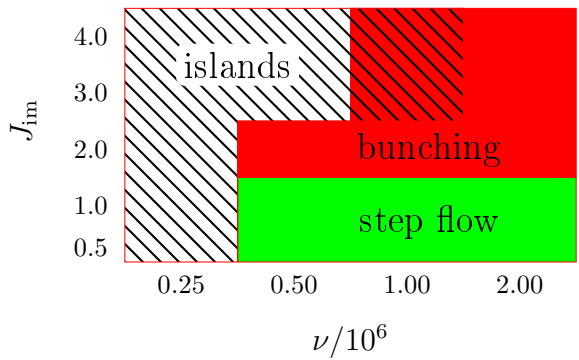} 
\caption{Summary of findings for the dominant growth mechanism in a 
  system with two steps as a function of $\Jimp$ and $\growth$.  
  The other system parameters are fixed to the values 
  $L_i=25$; $L_j=100$; $J=40$; $\Rimp = 0.1$.  
\label{fig:2stepSummary}} 
\end{figure}

\begin{figure} 
\includegraphics[width=0.5\textwidth]{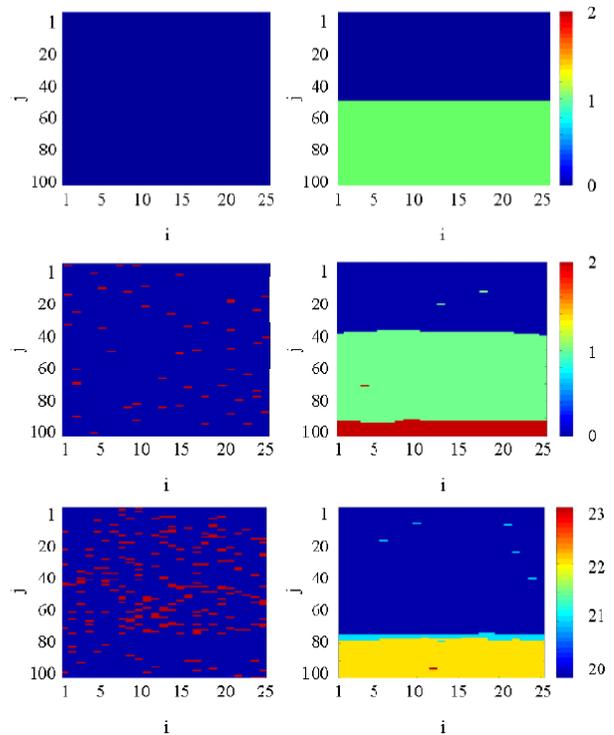} 
\caption{Impurity distributions (left) and color coded surface height 
  profiles (right) at three different growth stages for a system with 
  $\nrSteps=2$ steps which show step-pairing. The system parameters 
  are    
  $\Rimp=0.1$; $L_i=25$; $L_j=100$; $J=40$; $\Jimp=4$; 
  $\growth=2 \times 10^6$.  
  In the left panels red indicates impurities and blue normal atoms in 
  the topmost layer of the SOS representation of the 
  lattice. The respective heights are given in the right panels with a 
  color coding indicated by the bars on the far right.  
\\ 
  The \textbf{upper} panels show the initial configuration with two 
  equidistant straight steps, and no impurities on the surface.  
\\ 
  The \textbf{middle} panels show the situation after the deposition 
  of $0.2$~ML. The impurities are roughly uniformly distributed, and 
  the distances between the steps does not yet deviate much from the 
  initial configuration. 
\\%
  The \textbf{lower} panel shows the surface morphology after 
  deposition of $20$~ML, where the steps sit right next to each 
  other. Note that impurities at the terrace are not uniformly 
  distributed any more. There are more impurities ahead of the steps 
  than behind the steps. 
  \label{fig:2stepGrowth}} 
\end{figure}

Altogether the dynamics of surface growth is hence characterized by 
four dimensionless parameters summarized in \Tab{parameter}. Kinetic 
Monte Carlo simulation of this model (\cf\cite{Hegedus06} for details 
on the algorithm) show that the model is capable of reproducing the 
crossover from step flow at large $\growth$, where adatoms mostly 
attach to step edges, to island nucleation for smaller $\growth$, 
where adatoms merge into islands before reaching the step edges. 
Moreover, the simulations also show the formation of step pairs for 
appropriately chosen binding strength $\Jimp$ and density of 
impurities $\Rimp$ (see Sec.\ref{Sec:Dyn}). The different growth 
regimes for a system with $\nrSteps=2$ steps are summarized in 
\Fig{2stepSummary}. 
\Fig{2stepGrowth} shows snapshots of the time evolution of the surface 
height and the distribution of impurities for a system of two 
equidistant steps which evolves into a steady state where the two 
steps form a pair.

\section{Distribution of impurities} 
\label{Sec:Imp} 
 
Before further discussing the numerical results it is illuminating to 
calculate the distribution of impurities on the terraces. To this end 
we consider two steps which are roughly aligned in parallel such that 
they enclose a terrace of width $w$.  In a system of two parallelly 
aligned steps the width $w$ of the terrace does not change in a steady 
state where the steps form a bound pair, such as in the lower-most 
panels of \Fig{2stepGrowth}.  A small surface element is created in 
this situation when the edge of the step is formed at that position 
and height, and it is buried after deposition of two ML, when the two 
steps have reached the same position again. In order to study the 
distribution of impurities we use a comoving coordinate frame where 
$i$ denotes the position in lateral direction, and $\xj$ the 
distance from the position of the steps measured in the direction of 
growth.  
The latter distance will be measured in units of $L_j$, such that 
  there are $\xj \, L_j$ lattice sites between the considered site and 
  the step edge.
The age $\tau(\xj)$ of a position on the surface is 
proportional to $\xj L_j$. The age will be measured in units of atoms 
added to the surface since the site has last been visited by a step. 
For $\nrSteps=2$ bunched steps moving together it amounts to 
$\tau(\xj) = \nrSteps L_i L_j \, \xj$. 
 
\begin{figure} 
\hfill\includegraphics{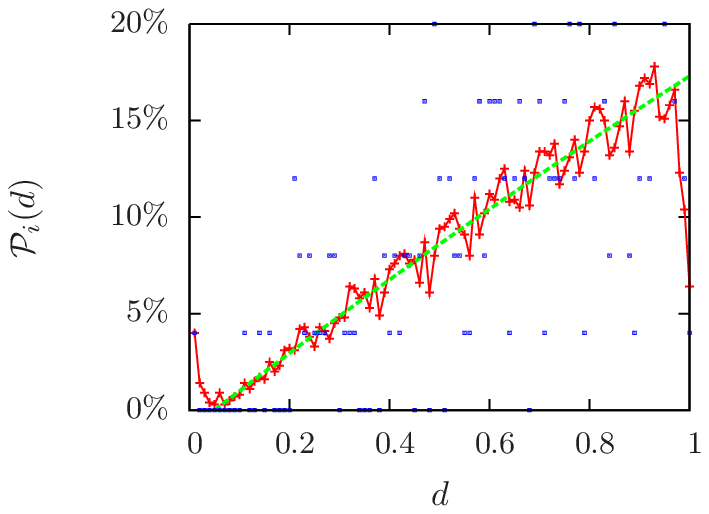}\rule{10mm}{0mm} 
\\[4mm] 
\hfill\includegraphics{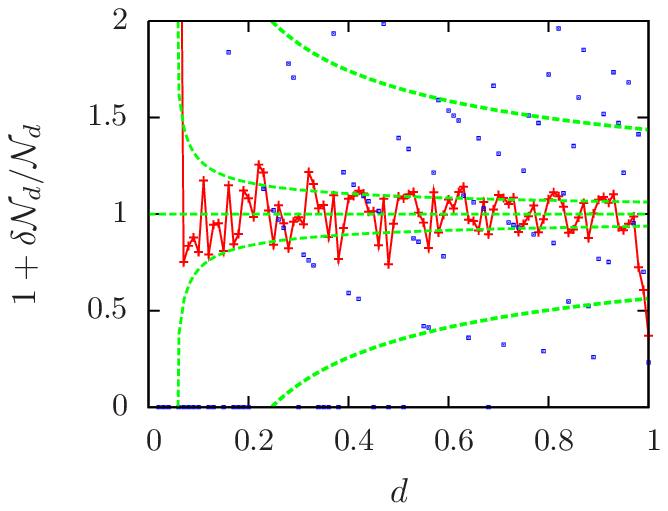}\rule{10mm}{0mm} 
\caption{Time average for the number of impurities as a function of the 
  distance $\xj$ from the step for the simulation shown in 
  \Fig{2stepGrowth}.  
  The upper panel shows a time average over a time span needed to grow 
  $8$~ML with numerical values indicated by red crosses and the theoretical 
  prediction \eq{densDist} by a dashed green line.  
  The lower panel shows the ratio of the numerical values and the 
  prediction (red crosses) together with those calculated from three 
  different snapshots. The dashed green lines give an estimate of the 
  width of the distribution for individual snapshots (thick line) and 
  a narrower band with a width that is smaller by a factor of five. 
  \label{fig:impDistribution}} 
\end{figure}

The probability that there is no impurity at that site can be 
calculated as follows: The probability to turn a site into an impurity 
when a single atom is added to the surface is  
$\probSingle = \Rimp / L_i L_j$.   
Hence, the probability to be changed to an impurity after $\tau(\xj)$ 
atoms have been added to the surface is 
\begin{eqnarray} 
\Prob_i (\xj)  
 & = & 
   1 - (1 - \probSingle)^{\tau(\xj)}  
\nonumber \\ 
 & = &  
   1 - \left( 1 - \frac{\Rimp}{L_i L_j} \right)^{ 
        {\nrSteps \Rimp \, \xj} \;  
        \frac{L_i L_j}{\Rimp} 
      } 
\nonumber \\[2mm] 
  & \simeq & 
    1 - \exp\left( - { \nrSteps \Rimp \, \xj} \right)  
\, . 
\label{eq:densDist} 
\end{eqnarray} 
The latter approximation applies provided that $L_i L_j / \Rimp \gg 1$. 
For the simulations shown in \Fig{2stepGrowth} one has  
$L_i L_j / \Rimp = 25\,000$  
such that this approximation is well justified. Moreover, in this 
situation the argument of the exponential function is small, such that 
the distribution of impurities is approximately linear, 
\begin{equation} 
\Prob_i (\xj)  
\equiv \nrSteps \Rimp \, \xj 
\, , 
\qquad \textrm{for \ } \nrSteps \Rimp \,\xj \ll 1  
\, . 
\label{eq:fewImps} 
\end{equation} 
\Fig{impDistribution} demonstrates that this prediction holds to a 
very good approximation for the time average over $8$~ML.  Plotting 
the ratio of the numerical values and the theoretical prediction 
(lower panel) shows that the agreement is better than $5$\% along the 
full width of the terrace. 
 
In contrast to time averages the distribution is very noisy, however, 
for snapshots of the distribution of impurities as shown by three 
different examples in the lower panel of \Fig{impDistribution}. 
Indeed, for the considered deposition process the number of impurities 
in a row $\xj$ is distributed according to a binomial distribution 
such that one expects to find  
$\mathcal{\nrSteps}_{\xj} = L_i \Prob_i (\xj)$  
impurities in a row at a normalized distance $\xj$ from the 
steps. The standard deviation should be  
$L_i \Prob_i (\xj) \left( 1 - \Prob_i (\xj) \right)$  
such that the relative error, which is indicated by a thick dashed 
green line in the lower panel of \Fig{impDistribution}, takes the 
value 
\[ 
\frac{\delta \mathcal{\nrSteps}_{\xj}}{\mathcal{\nrSteps}_{\xj}} 
= 
\left( 
  \frac{  1 - \Prob_i (\xj) }{  L_i \Prob_i (\xj) } 
\right)^{1/2}.  
\] 
For small \xj\ hardly any atoms impinged on that part of the surface 
such that the probability $\Prob_i$ is small. In that case the 
relative error exceeds $100$\%. From the perspective of understanding 
the transport problem this is not so problematic, however, because the 
impurities hardly influence the diffusion on the surface in that case. 
It is much more important to observe that the relative error remains 
fairly large even far away from the steps, where the coverage of 
impurities is large. In the case of \Fig{impDistribution} where we 
consider expectation values for $L_i=25$ and the final coverage with 
impurities is close to $25$\% the relative error still amounts to $( 
0.8 / (25 / 5) )^{1/2} = \sqrt{0.16} = 0.4$, and even for a coverage 
of $50$\% it only drops down to $0.2$. 
Fluctuations of this magnitude are troublesome, when trying to 
describe the transport by a continuum model which does not take into 
account fluctuations of the distribution of impurities.  The problem 
can not be resolved by considering averages of larger $L_i$ because 
the continuum description has to be based on local averages, and a 
one-dimensional description will only apply when the two-dimensional 
equations obtained in this manner are invariant under translation 
parallel to the steps. It is this latter property, however, which is 
lost when the fluctuations in the density of impurities are 
noticeable. 
 
The observation that the variance in the number of impurities in small 
neighborhoods of the lattice is large probably applies in general. 
This poses a major challenge to continuum models of step bunching 
where the presence of impurities is only taken into account as a 
modification of the diffusion coefficient, which itself depends on the 
expectation value for the number of impurities. In view of the large 
fluctuations of the distribution this might very well be a 
non-admissible oversimplification, which deserves a close inspection 
by comparison to numerical results in the following section. 
 
\section{Impurity-induced step dynamics} 
 
\label{Sec:Dyn} 
 
\subsection{Diffusion bias} 
 
The key ingredient of the continuum theory developed in \cite{Krug02} is the  
dependence of the effective adatom diffusion coefficient $D(\theta)$ on the  
local impurity coverage $\theta$.  
As we are concerned here 
with impurity concentrations of $\theta \approx 0.1$ or less,   
the leading term in an expansion in $\theta$ is expected to suffice. 
We therefore write 
\begin{equation} 
\label{eq:Dimp} 
D(\theta) \approx D(0) \; (1 - \alpha \theta).   
\end{equation} 
For completely blocking barriers ($\Jimp \to \infty$) the coefficient 
is given by $\alpha = \pi - 1 \approx 2.14$ \cite{Ernst87}.  
In effective medium approximation \cite{Haus87} one obtains the simple 
expression 
\begin{equation} 
\label{eq:alpha} 
\alpha = 2\; \frac{\Jimp - 1}{\Jimp + 1}. 
\end{equation} 
For $\Jimp \to \infty$ it yields $\alpha = 2$, 
which is close to the exact result.  
 
Using the general formulae derived in \cite{Krug02}, the adatom  
currents $j_\pm$ to the ascending ($j_+$) and descending ($j_-$)  
steps bordering a terrace of width $w$ can be computed from 
\Eq{Dimp}. For perfectly absorbing steps one finds that 
$j_\pm = \Phi w p_\pm$, where the attachment probabilities $p_\pm$ 
are independent of the terrace width $w$, and given by  
\begin{equation} 
   p_-  
   =  \frac{1}{\alpha\rho} + \frac{1}{\ln(1 - \alpha\rho)}  
\equiv 1 - p_+ 
\end{equation} 
where the relation $p_+ \equiv 1 - p_-$ ensures flux conservation.  
For small $\alpha\rho$ the attachment probabilities $p_\pm$ amount to 
\begin{equation} 
\label{ppm} 
   p_\pm 
   \approx \frac{1}{2} \mp \frac{\alpha}{12} \rho \, , 
\end{equation} 
such that the effect of the impurities can be quantified by the 
diffusion bias parameter 
\begin{equation} 
\label{eq:b} 
   b = p_- - p_+ \approx \frac{\alpha \rho}{6}. 
\end{equation} 
The sign is chosen such that step bunching results for $b > 0$ 
\cite{Krug02,Krug05}. 
 
\begin{figure} 
\[\rule{-5mm}{0mm} 
\includegraphics{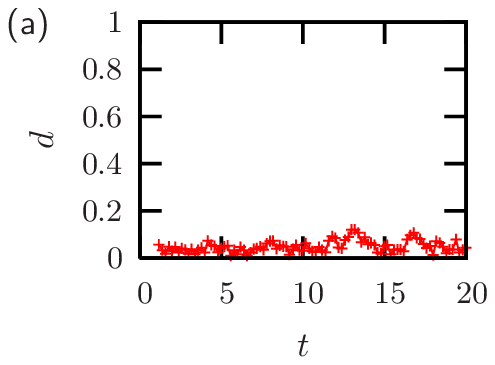} \quad  
\includegraphics{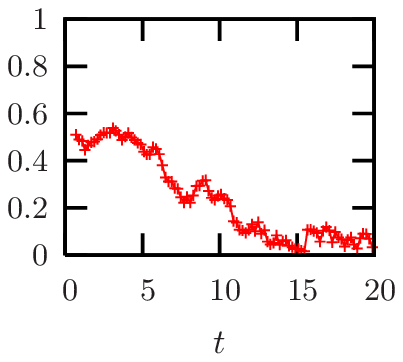} 
\]\[\rule{-5mm}{0mm} 
\includegraphics{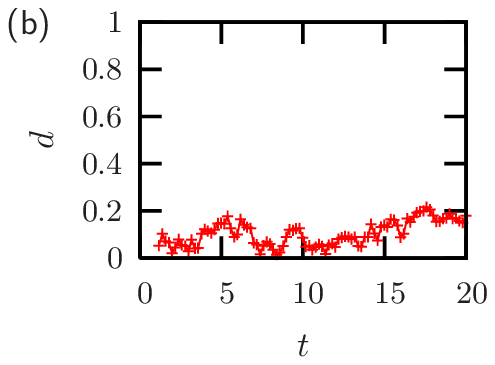} \quad  
\includegraphics{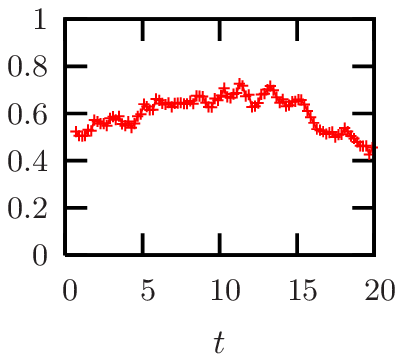} 
\]\[\rule{-5mm}{0mm} 
\includegraphics{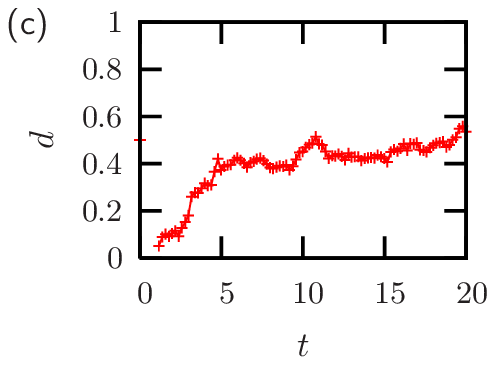} \quad  
\includegraphics{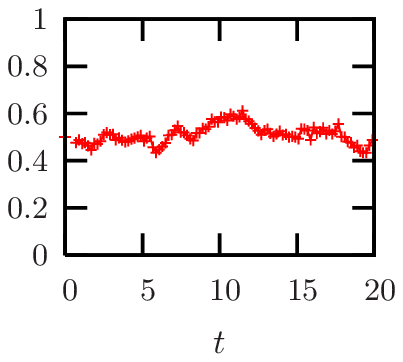} 
\] 
\caption{Time evolution of the normalized 
  step distance
  \dj\ for $\nrSteps=2$ steps and different choices of \Jimp:  
  (a) attractive, $\Jimp=4$,  
  (b) marginally stable, $\Jimp=1$, and 
  (c) repulsive, $\Jimp=0.2$, interaction between the steps. 
  The panels on the left show the evolution of an initial condition 
  where the two steps are right next to each other, and the right 
  panels that of an initial condition with equidistant steps.  
  The other system parameters were fixed to the values 
  $L_i=25$; $L_j/\nrSteps=50$; $J=40$; $\growth=2 \times 10^6$; $\Rimp=0.1$.  
  \label{fig:2stepTerraces}} 
\end{figure} 
 
The following considerations will be based on a simple 
one-dimensional, deterministic dynamical model for the step positions 
$x_k$, $k = 1,..,\nrSteps$ measured along the $j$-direction. The speed 
of the $k^{\mathrm{th}}$ step is the sum of the fraction $p_+$ of the 
flux incident on the (leading) terrace in front of the step, of width 
$x_{k+1} - x_k$, and the fraction $p_-$ of the flux incident on the 
(trailing) terrace behind the step, of width $x_k - x_{k-1}$. Thus we 
have 
\begin{equation} 
\label{evol} 
\frac{\rmd x_k}{\rmd t}  
= \frac{1}{2} (1 - b) (x_{k+1} - x_k) 
+ \frac{1}{2} (1 + b) (x_{k} - x_{k-1}) 
\end{equation} 
with the additional constraint that $x_k > x_{k-1}$ at all times.  
For convenience we use here dimensionless units where time is measured 
in units of the time scale   
$\inFlux^{-1}$ needed to deposit a monolayer of new 
material, and length still in units of $L_j$.

\subsection{Step-step interactions} 
 
Before turning to the discussion of our simulation results,  
we need to address the role of repulsive 
step-step interactions that are usually added to the right hand side 
of (\ref{evol}) \cite{Popkov06,Popkov05}. As no direct  
step-step interactions are included in our KMC model, we only   
consider the well-known entropic interactions induced by collisions 
between neighboring steps, which in turn are a consequence of thermal 
step meandering. Following \cite{Krug05}, we estimate 
the distance between two such collisions along the transverse 
($i$-) direction to be of the order of  
\begin{equation} 
\label{collision} 
L_c \sim \frac{\tilde \delta w^2}{\kB T}, 
\end{equation} 
where $\tilde \delta$ is the step stiffness and $w$
denotes the distance 
between the two steps. Clearly step collisions are irrelevant as long as 
$L_c$ is larger than the lattice size $L_i$ parallel to the steps.  
We therefore conclude that the range of the  
repulsive step-step interactions in our simulations is limited 
to step distances smaller than $w_c \sim (L_i \kB T/\tilde \delta)^{1/2}$. 
Using the expression \cite{Michely04,Krug05} 
\begin{equation} 
\label{stiff} 
\tilde \delta \approx \frac{\kB T}{2} \sqrt{J} 
\end{equation} 
for the step stiffness in the SOS model at low temperatures, we find 
that $w_c \approx 2.8$ for $L_i = 25$ and $J = 40$.  Thus the 
step-step interactions in our simulations are a purely local effect 
which merely ensures that steps cannot overtake each other.  In this 
sense the situation is similar to that considered in \cite{Slanina05} 
within a one-dimensional model with hard core step-step interactions. 
 
\begin{figure} 
 \includegraphics{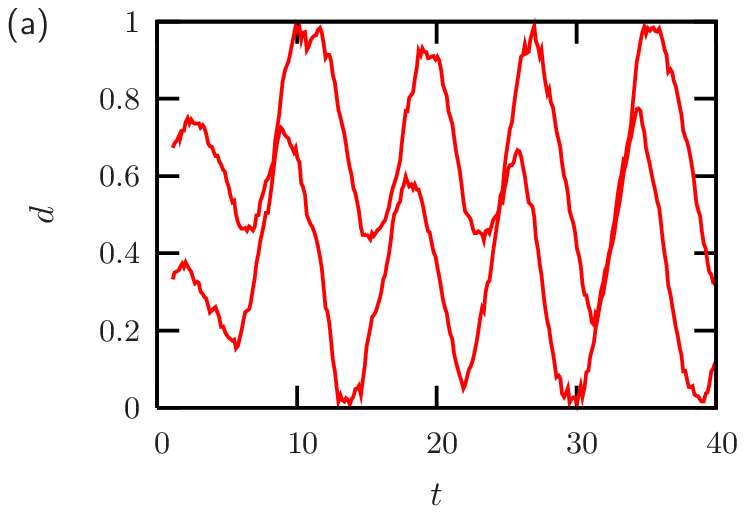}  
 \includegraphics{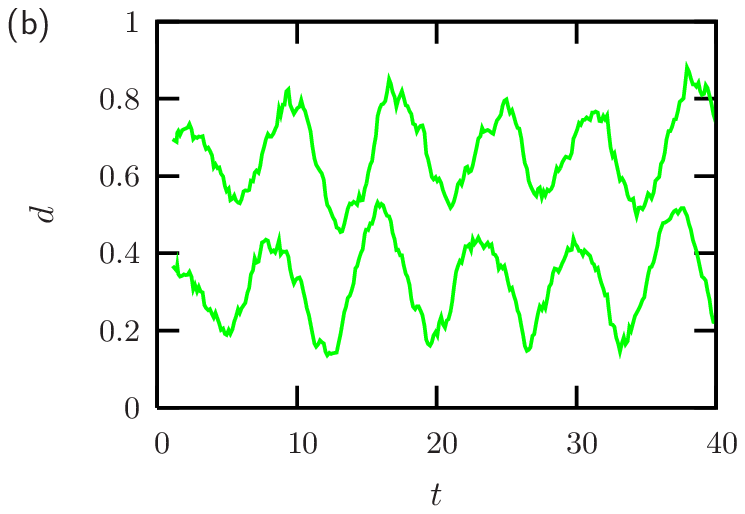}  
 \includegraphics{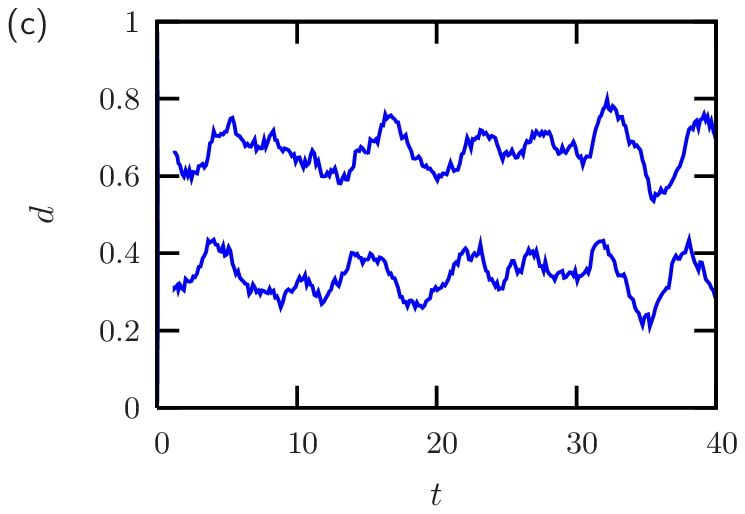}  
\vspace{5mm} 
\caption{Time evolution of normalized 
step distances $\dj_2$ and 
  $\dj_3$ for $\nrSteps=3$ steps and different choices of  
  \Jimp:   
  (a) attractive, $\Jimp=4$,  
  (b) marginally stable, $\Jimp=1$, and 
  (c) repulsive, $\Jimp=0.1$ interaction between the steps.  
  The other parameters are 
  $L_i=25$; $L_j/\nrSteps=50$; $J=40$; $\growth=3 \times 10^6$; $\Rimp=0.1$.  
  When impurities induce repulsion (c), the steps remain 
  well-separated, just as in the simulations with only two steps. 
  However, for three steps a smaller value of \Jimp\ ($\Jimp=0.1$ 
  rather than $\Jimp=0.2$ displayed in \Fig{2stepTerraces} for 
  two steps) was needed to clearly show this effect. 
  In the marginally stable case (b) the distances fluctuate showing 
  regularly looking oscillations. They arise due to coupling of step 
  velocities because of shared neighboring terraces. In this case the 
  steps approach each other much closer than in the repulsive case.  
  However, still they always remain well-separated due to entropic 
  repulsion, which is always present.  
  Finally, in the case of impurity-induced attraction between the 
  steps (a), one clearly sees bunches of two steps, which separate 
  however when the third step approaches.  
  \label{fig:3steps} } 
\end{figure} 
 
\subsection{Stability of step pairs}  
 
Consider first a system of two steps, with
normalized step distances $\dj(t)$ and $1-\dj(t)$ (\Fig{2stepTerraces}). 
It follows from (\ref{evol}) that $\dj(t)$ evolves according to  
\begin{equation} 
\label{l1} 
\dot{d} = b \; ( 2 d - 1 ). 
\end{equation} 
The equidistant fixed point $d = 1/2$ is unstable (stable) for $b > 0$ 
($b < 0$). For $b > 0$ the steps collide ($d \to 0$) in finite time, 
while for $b < 0$ a pair of close steps separates and approaches 
the fixed point exponentially at rate $2b$. 
Using the expression \eq{b} with $\alpha$ given by \Eq{alpha} the 
half-time of the decay into the stable states, $t_{1/2} \equiv \ln 2 / 
2b$ is about $16$~ML.  This value is consistent with but somewhat 
larger than the time scale observed in the simulations 
\Fig{2stepTerraces}(a) and (c).  A possible source of this deviation 
is the fact that the impurity concentration profile may not have 
reached stationarity on the time scale of step motion.

\begin{figure} 
\[ \includegraphics{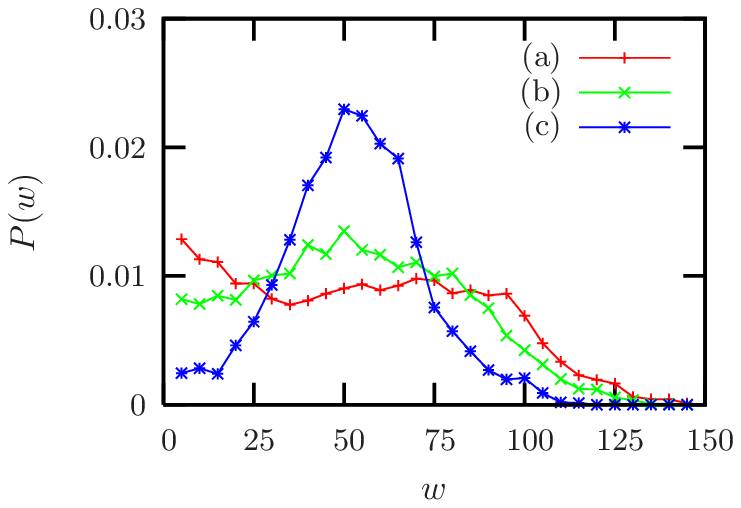} \] 
\[ \includegraphics{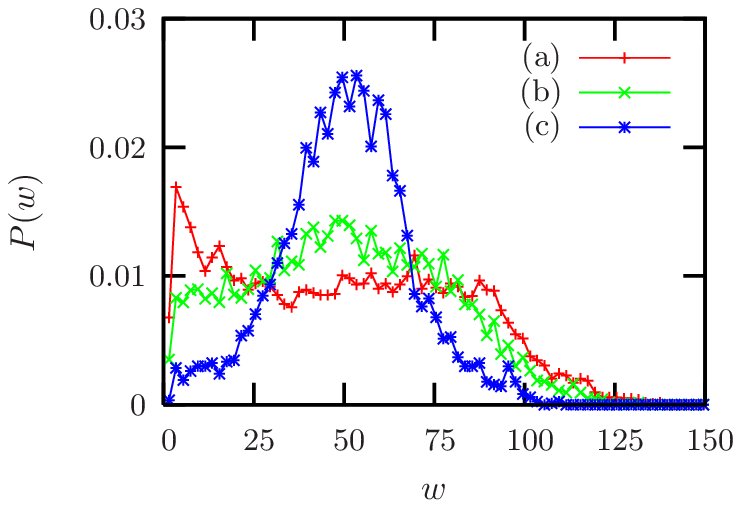} \] 
\caption{ 
%
  The probability densities to find two adjacent steps at a distance 
  $w$ for systems with $\nrSteps=3$ (top) and 
  $\nrSteps=8$ (bottom) steps, respectively.  For $\nrSteps=3$ the 
  lines (a)--(c) refer to the data shown in the respective panels of 
  \Fig{3steps}, and the simulation with $\nrSteps=8$ was run with  
  $L_i=20$; $L_j/\nrSteps=50$; $J=40$; $\growth=8 \times 10^6$; $\Rimp = 0.1$;  
  and (a) $\Jimp=4$, (b) $\Jimp=1$, (c) $\Jimp=0.2$.  
  The slightly smaller width $L_i=20$ and larger $\growth$ was chosen 
  to minimize the impact of island formation. As also suggested by the 
  histograms this change has no significant impact on the step 
  interaction.  
  \label{fig:hist}} 
\end{figure} 
 
\subsection{Stability of small bunches}  
 
Consider next a system of three steps, where $\dj_2$ and 
  $\dj_3$ denote the normalized distance from the first to the 
  second and the first to the third step, respectively.
\Fig{3steps} shows the evolution of $d_2$ and $d_3$
as a function of time $t$ measured in units of deposited ML.  
 
In the absence of impurity-induced step interactions ($\Jimp=1$, 
\Fig{3steps}(b)) and for repulsive impurity-induced step 
interactions ($\Jimp=0.2$, \Fig{3steps}(c)) the system behaves very 
similar to the one with only two steps. Indeed, histograms for the 
probability to find a certain distance $w$ between adjacent steps in 
\Fig{hist} consistently show sharply peaked distributions around the 
the average terrace size $w=50$, while the distribution is broad with 
a maximum at this value in the neutral case.  
As expected for a system with attractive interactions between the 
steps, the distribution $P(w)$ has considerably more weight for small 
$w$. However, surprisingly, it does not decay but saturates at fairly 
large constant background extending till $w=100$. 
 
The origin of this background becomes clear from inspection of the 
time traces of $d_2$ and $d_3$ shown in \Fig{3steps}(a). The 
simulation shows the transient formation of step pairs which exchange 
partners at regular intervals. However, no stable step triplets are 
formed. To see how this follows from the dynamical equations 
(\ref{evol}), suppose a triplet of three nearby steps has formed, such 
that the step positions satisfy $x_2 - x_1, \, x_3 - x_2 \ll 1$, \ie 
they are both much smaller than the system size $L_j$. Then step 3 has 
a large terrace of size $\approx 1$ in front of it, and it moves at 
speed $(1 - b)/2$. Step 2 is surrounded by small terraces and moves 
very slowly, and step 1 is constrained by the no-passing condition to 
move at the same speed. Thus for \emph{every} $b < 1$ step 3 will 
detach from the triplet.

\begin{figure} 
\includegraphics{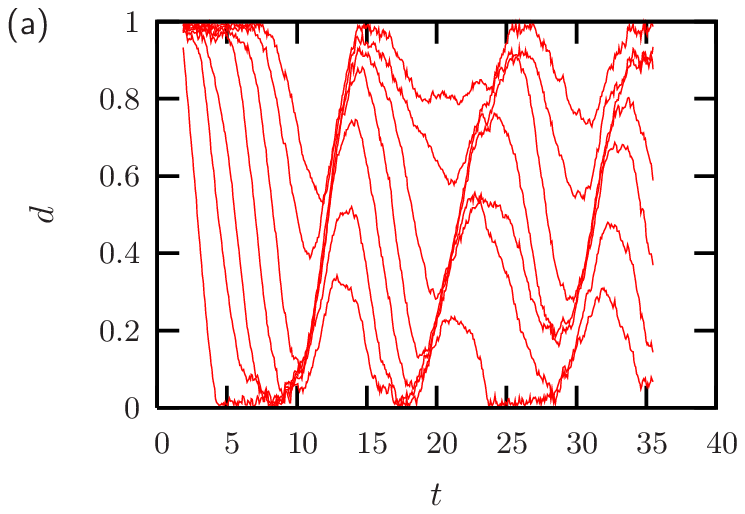} 
\includegraphics{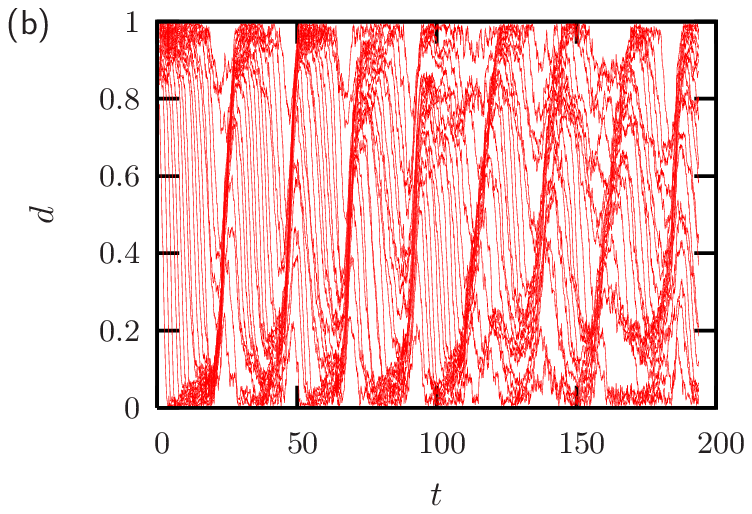} 
\caption{Evolution of a bunch of eight (top) and twenty (bottom) steps 
  in space-time representation.  
  The simulation with $\nrSteps=8$ was run  
  with the same parameters as in \Fig{hist} (bottom,a) except for 
  a larger value $\growth=2.4 \times 10^7$, and the one with $\nrSteps=20$  
  with $L_i=20$; $L_j/\nrSteps=50$; $J=40$; $\Jimp=4$; 
  $\growth=3 \times 10^7$ and $\Rimp=0.1$.  
  \label{fig:decay}} 
\end{figure} 
 
We conclude that, for any $b < 1$, step triplets and larger bunches 
are unstable against the emission of single steps.  Bound states of 
steps moving together at constant speed \cite{Sato97} can form when $b 
> 1$ \cite{Popkov06}, but this condition obviously cannot be reached 
in the growth model considered here.  A detailed analysis of the 
equations (\ref{evol}) shows that step emission will continue until 
the number of free steps between two bunches (or, equivalently, 
between one bunch and its periodic image) has reached the steady state 
value \cite{Popkov06} 
\begin{equation} 
\label{N_f} 
\nrSteps_f \approx \frac{1}{3b} \ln \nrSteps, 
\end{equation} 
where \nrSteps\ denotes the total number of steps in the bunch and on 
the terrace. As $\nrSteps_f$ cannot exceed \nrSteps, we 
conclude that a stable bunch can form only if the number of steps 
satisfies the condition  
\begin{equation} 
\label{bound} 
3 b \, \nrSteps / \ln \nrSteps > 1.  
\end{equation} 
With the value of $b \approx 0.02$ obtained for $\Jimp = 4$ and $\rho = 0.1$ 
this implies $\nrSteps > 65$, which (given the constraints on the 
systems parameters described above) exceeds our computational 
capacities.  
Indeed, simulations conducted with systems containing up to 20 steps 
confirm that step bunching remains a transient phenomenon, even when a 
bunched initial configuration is chosen (\Fig{decay}). 
 
\begin{figure} 
\[ 
\includegraphics[width=0.7\linewidth]{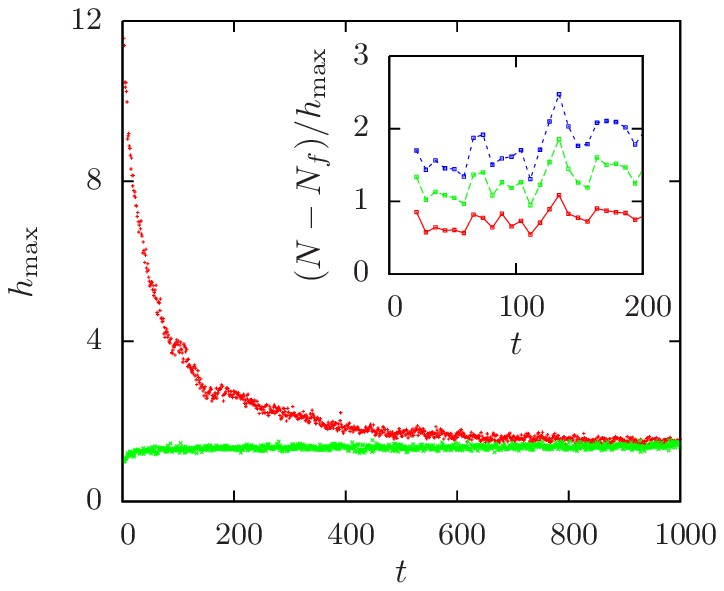} 
\]\[ 
\includegraphics[width=0.7\linewidth]{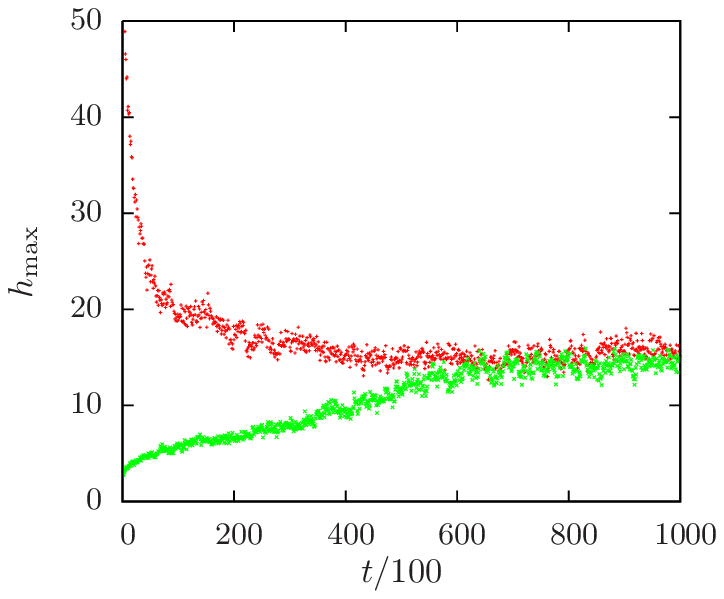} 
\]\[ 
\includegraphics[width=0.7\linewidth]{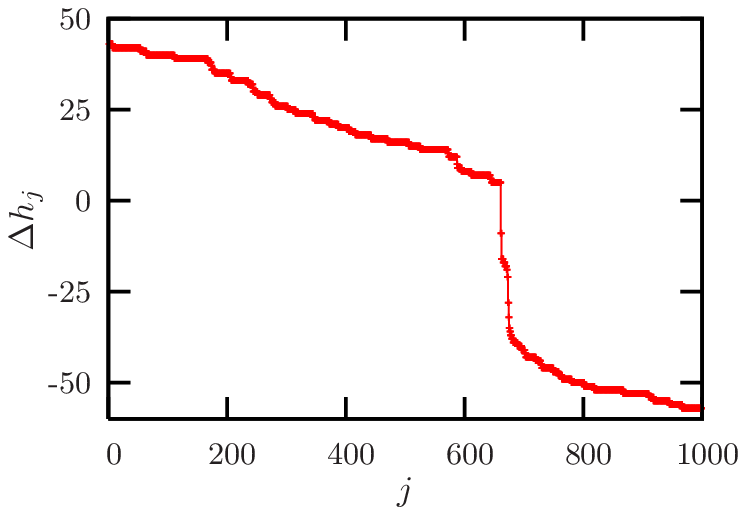} 
\] 
\caption{Simulations of step bunching in a one-dimensional model. 
The top and middle panels show the time evolution of the maximal step 
height for systems containing $\nrSteps = 20$ and \nrSteps = $100$ 
steps, respectively. The respective data points in these plots 
correspond to initial conditions with a single large step (red) and an 
equidistant step train (green), which were averaged over $100$ 
independent runs. Beware of the different time scales of these two 
plots. 
The inset in the upper graph shows the ratio of the data of 
the 1D model (red curve) to the maximal height of a bunch in the 
corresponding KMC simulation \Fig{decay}. The bunch size was 
calculated by determining the maximal number of steps in a window of 
fixed width $d=0.03$, $d=0.06$, and $d=0.09$ (from bottom to top) 
roughly corresponding to the range of the effective hard-core repulsion 
$\nrSteps w_c/L_j$.
The lowermost panel shows the deviation $\Delta h_j \equiv 
h_j-10^5$ of the height $h_j$ from the average height after deposition 
of $10^5$~ML. It is the final configuration of one of the 
simulations run to generate the graph in the middle panel. 
\label{fig:1Dmodel}} 
\end{figure} 

\subsection{Stability of large bunches}  
 
The agreement between the predictions and the results of the kinetic 
Monte-Carlo simulations indicates that the stability of bunches 
can be captured based on the one-dimensional growth model.  To 
illustrate the formation of stable bunches with increasing \nrSteps\ 
we therefore relied on simulations of a one-dimensional stochastic 
growth model described in detail in \cite{Slanina05}. 
In this model particles are deposited onto a one-dimensional vicinal 
surface and transferred instantaneously (without explicit diffusion) 
to the ascending or descending step with probabilities $p_+$ and 
$p_-$, respectively. Since steps are allowed to coalesce (but not to 
pass each other), the formation of bunches can easily be followed by 
monitoring the maximal step height $h_{\mathrm{max}}$ (the largest 
nearest neighbor height difference) in the system. In \Fig{1Dmodel} 
we show results obtained for $b = 0.02$ on a lattice of $L_j = 1000$ 
sites. For $\nrSteps = 20$, the situation corresponding to 
\Fig{decay}(b), the maximal step height remains below 2, showing that 
only step pairs exist, whereas for $\nrSteps = 100$ a stable bunch 
forms that contains almost half of the steps in the system.

\section{Conclusions} 
\label{Sec:Con}

We summarize the main achievements of this work:  
 
(i) We have verified by KMC simulations that an impurity-induced 
increase of the transition energy between neighboring sites, a purely 
kinetic effect, is a possible source of step bunching. By focusing on 
the behavior of pairs of steps we have explicitly determined the sign 
and strength of the impurity-induced step interactions, as quantified 
by the asymmetry parameter $b$.  
The adatoms can make use of massive fluctuations of the spatial and 
temporal distribution of the impurities (\Fig{impDistribution}) in 
order to find optimal paths to the binding sites at the steps. 
As a consequence an estimate of the order of magnitude of $b$ 
properly has to account for the diffusion of the adatoms in a 2d 
disordered arrangement of impurities, which is very different from the 
1d setting used in \cite{Krug02}. 
 
(ii) The step bunching observed in our work is substantially weaker 
than that found in simulations of the SiC system \cite{Croke00}, where 
the impurities were assumed to affect the adatom binding energies. 
This probably indicates that the effective asymmetry $b$ is larger for 
energetic impurities. Unfortunately, although the theory of 
\cite{Krug02} correctly predicts step bunching for impurities that 
\textit{lower} the binding energy, the magnitude of the effect depends 
on the precise boundary conditions at the steps, which do not easily 
translate into the two-dimensional KMC setting. 
 
(iii) Despite the presence of impurity-induced attractive step 
interactions, triplets and larger assemblies of steps are not 
necessarily stable when $b$ is small: Instead of agglomerating into 
macroscopic step bunches, the steps display a peculiar dynamical 
pattern characterized by transient step pairs that exchange partners 
much like in a folk dance.  
Up to now this effect has gone unnoticed, because previous KMC 
simulations of step bunching during growth have generally considered 
situations where the effective attachment asymmetry $b$ is of order 
unity \cite{Sato2001,Xie2002,Videcoq2002}. 
The lack of stability of the step assemblies was explained based on a 
recently developed determinstic theory \cite{Popkov06}. It allows us 
to predict that bunches can form only when the number of steps exceeds 
the bound (\ref{bound}). 
 
(iv) For real surfaces there is no restriction on the total number of 
steps. Nevertheless, it is highly improbable to observe bunching in 
systems with small $b$. Our simulations for the one-dimensional model 
(\Fig{1Dmodel}) show that for small $b$ bunches evolve only after 
exceedingly long times even when the bound (\ref{bound}) is satisfied: 
In order to see step bunching one has to wait for a fluctuation 
nucleating a bunch with a minimal size given by (\ref{bound}). 
 
For applications the most noticeable consequence of our study is that 
the rapid formation of large step bunches seen experimentally in the 
growth of SiC \cite{Croke00} cannot be explained only in terms of 
kinetic impurities. Some coupling to the adatom binding energy must 
also be involved.

\acknowledgments 
 
J.V.\ and J.H.\ are grateful to Peter Thomas for support and steady 
encouragement, and acknowlege illuminating discussions with Peter 
Jacob and coworkers on steps and their dynamics.  J.K. acknowledges 
useful discussions with Vladislav Popkov, as well as the kind 
hospitality of MPI-DS (G\"ottingen) and LPT/ENS (Paris) where 
part of the paper was written. 
\\ 
The project has been supported by the DFG.


 

\end{document}